\documentclass[10pt, conference, letterpaper]{ IEEEtran}
\usepackage{wrapfig}
\usepackage{cite}
\usepackage{verbatim}
\usepackage{enumitem} %
\usepackage[T1]{fontenc}
\usepackage{tablefootnote}
\usepackage{amsfonts}  
\usepackage[table]{xcolor}
\usepackage{wrapfig,lipsum}
\renewcommand{\baselinestretch}{.9465}
 \usepackage{tikz}
\usetikzlibrary{shapes, arrows, positioning, fit}
\usepackage{booktabs} 
\usepackage{siunitx}  
\usepackage{multirow} 
\usepackage{multirow}
\usepackage{graphicx}
\usepackage{array}

\usepackage{float}
\usepackage{caption}
\usepackage{wrapfig}
\usepackage{tabularray}
\usepackage{amsmath}
\usepackage{amssymb}
\usepackage{threeparttable}
\usepackage{booktabs}
\usepackage{multirow}
\usepackage{multicol}
\usepackage{tabularx}
\usepackage{subcaption}
\usepackage[margin=0.68in]{geometry}
\IEEEoverridecommandlockouts
\usepackage{enumitem}

\usepackage[linesnumbered,ruled,vlined]{algorithm2e}
\usepackage{xcolor}
\usepackage{tabularx}
\usepackage{colortbl}
\usepackage{orcidlink}

\definecolor{lime}{rgb}{0.88,2,10}

\usepackage[spaces,hyphens]{xurl}
\newcommand*{\Resize}[2]{\resizebox{#1}{!}{$#2$}}%
\usepackage{amssymb}
\usepackage{textcomp}

\newcommand{\fref}[1]{Fig.~\ref{#1}}

\newcommand{\sref}[1]{Section~\ref{#1}}
\usepackage{fancyhdr}
\fancypagestyle{mystyle}{
    \chead{\large The 13th IEEE International Conference on Big Data (IEEE BigData 2025)}}

\author{
\IEEEauthorblockN{
Mohamed Elmahallawy\IEEEauthorrefmark{1},
Asma Jodeiri Akbarfam}
\IEEEauthorblockA{
School of Engineering \& Applied Sciences, Washington State University, Richland, WA 99354, USA\\
Emails: \{mohamed.elmahallawy, asma.akbarfam\}@wsu.edu}
\IEEEcompsocitemizethanks{
\IEEEcompsocthanksitem \IEEEauthorrefmark{1}Corresponding author.
}\vspace{-0.5cm}
}
\newcommand\HUGE{\fontsize{20.5}{25}\selectfont}
\begin{document}

\title{\HUGE{Decentralized Trust for Space AI: Blockchain-Based Federated Learning Across Multi-Vendor LEO Satellite Networks}}

\maketitle
\thispagestyle{mystyle}
 
\begin{abstract}

The rise of \textbf{\em space AI} is reshaping government and industry through applications such as disaster detection, border surveillance, and climate monitoring, powered by massive data from commercial and governmental low Earth orbit (LEO) satellites. \textit{Federated satellite learning (FSL)} enables joint model training without sharing raw data, but suffers from slow convergence due to intermittent connectivity and introduces critical \textit{trust} challenges—where biased or falsified updates can arise across satellite constellations, including those injected through cyberattacks on inter-satellite or satellite–ground communication links. We propose \textbf{\textsc{OrbitChain}}, a blockchain-backed framework that empowers trustworthy \emph{multi-vendor} collaboration in LEO networks. \textsc{OrbitChain} (i) offloads consensus to high-altitude platforms (HAPs) with greater computational capacity, (ii) ensures transparent, auditable provenance of model updates from different orbits owned by different vendors, and (iii) prevents manipulated or incomplete contributions from affecting global FSL model aggregation. Extensive simulations show that \textsc{OrbitChain} reduces computational and communication overhead while improving \emph{privacy, security, and global model accuracy}. Its permissioned proof-of-authority ledger finalizes over 1{,}000 blocks with sub-second latency (0.16,s, 0.26,s, 0.35,s for 1-of-5, 3-of-5, and 5-of-5 quorums). Moreover, \textsc{OrbitChain} reduces convergence time by up to 30 hours on real satellite datasets compared to single-vendor, demonstrating its effectiveness for real-time, multi-vendor learning. Our code is available at \href{https://github.com/wsu-cyber-security-lab-ai/OrbitChain.git}{GitHub}


\end{abstract}


\section{Introduction}
Advances in artificial intelligence (AI) together with decreasing satellite launch costs have enabled dense low Earth orbit (LEO) constellations that generate massive, heterogeneous remote-sensing streams (up to 5~PB/day) from diverse sensors and orbits \cite{abdelsadek2022future,lucia2021computational,wang2022enhancing}. These constellations—deployed by commercial operators (e.g., SpaceX, Amazon, OneWeb) and government agencies (e.g., NASA, ESA)—power applications ranging from global Internet access to environmental monitoring, disaster response, border security, and smart-city services \cite{gupta2019xbd,perez2021airborne,routray2020military,luo2024leo}. The traditional approach—downloading each vendor's raw imagery to its ground station (GS) for local model training—fails to scale because LEO links are intermittent and short (typically a few daily passes of $\sim$5--10 minutes), bandwidth is constrained relative to the volume of high-resolution data, and raw data transfers raise acute privacy and security risks.\cite{leyva2023satellite}

Federated satellite learning (FSL) \cite{elmahallawy2024stitching} offers a compelling alternative: satellites act as space-edge clients that train local models on private data and share model updates to a parameter server $\mathcal{PS}$ rather than raw data. However, FSL across heterogeneous LEO constellations often suffers from poor convergence accuracy and long convergence times—driven by intermittent links, non-i.i.d. data distributions, device and sensor heterogeneity, and stale or asynchronous updates \cite{madoery2024novel, mAsyFLEO}. Encouraging {\em ``multi-vendor''} collaboration provides a practical remedy: by aggregating updates from diverse constellations, each vendor's model is exposed to wider environmental contexts, more instances of rare events, and complementary sensor modalities. This cross-vendor knowledge fusion reduces distributional bias, accelerates convergence, improves generalization, and increases robustness to sensor noise and occlusions—raising each vendor’s \emph{individual} operational accuracy without necessitating raw-data exchange.

Yet, realizing {\em multi-vendor collaboration} in FSL is far from trivial and introduces several technical, operational, and commercial challenges. First, \emph{trust and verification}: vendors must be able to verify the provenance and integrity of contributed updates while ensuring that no other vendor can access or infer their private data \cite{yue2023low,akbarfam2150sok}. Second, \emph{heterogeneity and interoperability}: satellites vary in communication windows, compute power, and memory capacity, which complicates model aggregation, introduces stragglers, and risks biasing the global model toward contributions from more capable vendors \cite{lin2024fedsn}. Third, \emph{incentive alignment and governance}: competing commercial interests, ambiguous intellectual property claims over model weights, and the risk of free-riding discourage honest participation unless transparent incentives and enforceable agreements are in place. Finally, \emph{collusion risk}: $\mathcal{PS}$s—such as GSs or high-altitude platforms (HAPs)\cite{elmahallawy2024communication}—owned by one vendor may collude with select vendors to bias the global model in their favor, undermining fairness and trust.



To address the aforementioned challenges, we propose {\em\textsc{OrbitChain}}, an innovative blockchain-based framework specifically designed for FSL to enforce trust, improve robustness, accelerate convergence, and enhance model accuracy in multi-vendor environments. \textsc{OrbitChain} leverages a blockchain ledger as a transparent, tamper-evident record of all FSL updates. After each training round, the aggregated global model—or its cryptographic hash, along with metadata such as the round number, contributing satellite nodes, and satellite identifier—is committed to the blockchain. This design produces an immutable audit trail that records how each global model version was formed and which participants contributed to it, enabling verifiable provenance and accountability across vendors. Because the ledger is append-only, it prevents retroactive modification of historical models or contribution records, thereby strengthening trust among competitive stakeholders and supporting post-hoc forensic analysis of manipulated local or global models. By anchoring the entire FSL process to a distributed ledger, \textsc{OrbitChain} transforms model aggregation from an opaque operation into a fully auditable and secure pipeline tailored for multi-vendor satellite networks.

\noindent In summary, this paper makes the following key contributions:

\begin{itemize}[leftmargin=*]

\item We propose  {\textsc{OrbitChain}}, a blockchain-based framework for  FSL that enforces trust, enhances robustness, accelerates convergence, and improves model accuracy across {\em multi-vendor satellite} constellations. To the best of our knowledge, it is the first to support \textit{large-scale multi-vendor collaboration}, handling heterogeneous constellations, with a combination of various $\mathcal {PS}s$ (GSs and HAPs).

\item  \textsc{OrbitChain} develops mechanisms that: (i) provide a \textit{tamper-evident, transparent ledger} that records all local and global model updates—along with cryptographic hashes and structured metadata—to ensure verifiable provenance, accountability, and post-hoc forensic analysis; and (ii) \textit{prevent fraudulent or malicious model contributions} from being accepted into the ledger, thereby mitigating cyber threats while preserving privacy across multiple satellite vendors. This is enabled through a Proof-of-Authority (PoA) consensus layer specifically designed for resource-rich HAP validators, delivering deterministic block finality and high throughput.


\item Extensive evaluations show that \textsc{OrbitChain} enables secure, privacy-preserving, and auditable multi-vendor FSL in LEO networks, achieving up to 30 hours faster convergence, improved per-vendor accuracy and robustness on real-world satellite datasets, and reduced computational and communication overhead. Our Flask-based PoA prototype processed over 1{,}000 blocks with block-finalization latencies of $0.16$–$0.35$\,s across 1-of-5, 3-of-5, and 5-of-5 quorum settings, demonstrating low-latency consensus, tamper-evident model aggregation, and \textsc{OrbitChain}’s real-time feasibility for LEO FSL operations.

\end{itemize}

\section{Related Work}
\label{sec:related}

\noindent While there is a growing interest in harnessing FL for LEO satellites and edge-enabled space architectures \cite{chen2023edge,e2023opt,mAsyFLEO,wu2023fedgsm,chen2024hermes, elmahallawy2023one}, most prior efforts focus on communication-efficient training and convergence optimization and pay limited attention to the unique trust, incentive, and provenance requirements that arise when multiple commercial vendors—each operating their own constellations and $\mathcal{PS}s$ (e.g., GSs or HAPs)—must collaborate. Below, we summarize relevant threads of literature and highlight why enabling {\em multi-vendor} collaboration in FSL remains an open problem.

\subsection{Secure aggregation and privacy-preserving primitives}
Secure aggregation, homomorphic encryption (HE), multi-party computation (MPC), functional encryption (FE), and differential privacy (DP) are commonly proposed tools to protect client data and model updates in FL \cite{bonawitz2017practical,wu2023esafl,ma2022privacy,chen2024secure,qian2024decentralized,wei2020federated,chen2022fundamental,elmahallawy2025secure,elmahallawy2023secure}. HE/FE enable computation on encrypted updates but typically inflate ciphertext sizes and communication cost and often require complex key-management across intermittently connected satellites. MPC provides strong privacy guarantees but incurs heavy rounds of interaction and computation that are challenging under LEO visibility constraints. DP is lightweight but trades privacy for utility and needs careful calibration to avoid degrading convergence. While these primitives address \emph{individual} confidentiality and secure aggregation, they do not by themselves provide an auditable, tamper-evident record of \emph{who contributed what} or resolve commercial governance and questions that are central to multi-vendor collaboration.

\subsection{Blockchain and ledger-based designs for FSL}
Recent research has explored integrating blockchain technology with FL to enhance decentralization, auditability, and incentive alignment in terrestrial settings. In essence, blockchain integration mitigates reliance on a $\mathcal{PS}$ and fosters trust among distributed participants~\cite{yang2024blockchain}. Wang \textit{et al.}~\cite{wang2021blockchain} classify blockchain-based FL (BCFL) architectures into {\em three categories:} (i) \emph{Fully Coupled (FuC-BCFL)}~\cite{ramanan2020baffle}, where each FL client also functions as a blockchain node; (ii) \emph{Flexibly Coupled (FlC-BCFL)}~\cite{kim2019blockchain, zhao2020privacy}, in which FL and blockchain operate semi-independently; and  (iii) \emph{Loosely Coupled (LoC-BCFL)}~\cite{kang2020scalable, ur2020towards}, where blockchain primarily supports reputation, auditability, and incentive management. These frameworks strengthen coordination and trust among clients; however, they do not fully resolve fundamental FL challenges such as data heterogeneity (non-i.i.d distributions), privacy leakage, and secure aggregation~\cite{nicolazzo2024privacy, zhu2021federated, ma2022state}. Moreover, some BCFL designs assume that every client can participate as a validator—an impractical requirement in FSL due to the limited onboard LEO satellites' computational and energy resources. Others lack cryptographic mechanisms that ensure the committed model updates are verifiably reflected in the aggregated global model.

\subsection{Convergence and system-level solutions for FSL}
A number of works target LEO-specific convergence problems: leveraging HAPs to accelerate aggregation \cite{happaper,elmahallawy2024communication}, clustering clients by channel quality \cite{chen2023edge}, compensating gradient staleness using orbital topology \cite{wu2023fedgsm}, and grouping models from different orbits to mitigate bias \cite{mAsyFLEO}. These papers improve training efficiency and robustness to orbital dynamics, but they assume a cooperative federation under a trusted coordinator and do not address cross-vendor trust, model provenance, or incentive structures required when competing operators collaborate.

\subsection{Gaps: enabling multi-vendor collaboration in LEO}

To the best of our knowledge, there is no prior work that holistically targets \emph{multi-vendor} FL for LEO satellite networks while ensuring both verifiability and confidentiality. A comprehensive solution must combine:  (i) ledger-backed, auditable provenance of model versions and contributor identities across heterogeneous $\mathcal{PS}s$  deployments (GSs/HAPs); (ii) consensus and ledger designs tailored to intermittent LEO connectivity and validator constraints; and (iii) incentive and governance mechanisms that reconcile competing commercial interests while preserving confidentiality of raw data and vendor-sensitive model internals.

Although prior FL and secure aggregation frameworks offer valuable primitives, and blockchain-enabled FL demonstrates the advantages of on-chain auditability in terrestrial contexts, the intersection of intermittent connectivity, extreme heterogeneity, commercial competition, and jurisdictional constraints in multi-vendor LEO systems remains unexplored. This gap motivates \textsc{OrbitChain}—a ledger-anchored FSL architecture designed {\em to achieve provable model provenance, collusion resistance among participating $\mathcal{PS}$s, and lightweight consensus and data-anchoring mechanisms tailored for the operational realities of satellite ecosystems.}

\section{System and Threat Model}\label{sec:system_model}

We model a multi-vendor FSL system composed of heterogeneous LEO constellations, shared HAPs, and vendor-owned  GSs. Below, we define the notation and describe the learning/communication/ledger operations mathematically.

 \subsection{FSL Network and Satellite Training}
Let $\mathcal{V}=\{1,\dots,V\}$ denote the set of vendors. Each vendor $v \in \mathcal{V}$ owns a constellation $\mathcal{K}_v \in K_v$ LEO satellites; the full set of satellites is $\mathcal{K}=\bigcup_{v\in\mathcal{V}}\mathcal{K}_v,$ where $ K \triangleq |\mathcal{K}|$. Also, each vendor $v$ operates a set of GSs $\mathcal{G}_v$. In addition to vendor-owned assets, the system deploys a shared set of high-altitude platforms (HAPs) $\mathcal{H}=\{1,\dots,H\}$, which act as the \emph{in-space federated $\mathcal{PS}$} and \emph{blockchain ledger validators}. 
Each satellite $k \in \mathcal{K}_v$ collects a local dataset $\mathcal{D}_{v,k}=\{(x_{v,k,j},y_{v,k,j})\}_{j=1}^{|\mathcal{D}_{v,k}|}$ for a given ML task (e.g., wildfire detection, ship tracking). During communication round $t$, let $\mathcal{K}^t \subseteq \mathcal{K}$ denote the set of satellites that are \emph{visible} to at least one HAP (see Sec.~\ref{sec:sys_mdl_visibility}). The total size of data participating in round $t$ is $ |\mathcal{D}^{t}|=\sum_{v\in\mathcal{V}}\sum_{k\in\mathcal{K}_v} |\mathcal{D}_{v,k}| \cdot \mathbf{1}\{k \in \mathcal{K}^t\}$. Each satellite optimizes a local empirical risk
\begin{equation}
F_{v,k}(\boldsymbol{\theta}) \triangleq 
\frac{1}{|\mathcal{D}_{v,k}|}
\sum_{(x,y)\in\mathcal{D}_{v,k}} f(\boldsymbol{\theta};x,y),
\end{equation}
where $f(\boldsymbol{\theta};x,y)$ is the sample-wise loss. The global FSL objective is to minimize the weighted average of local losses:
\begin{equation}
F(\boldsymbol{\theta}) \triangleq 
\frac{1}{|\mathcal{D}^{t}|}
\sum_{v\in\mathcal{V}}
\sum_{k\in\mathcal{K}_v \cap \mathcal{K}^t}
|\mathcal{D}_{v,k}|\,F_{v,k}(\boldsymbol{\theta}).
\end{equation}

At round $t$, the current global model is $\boldsymbol{\theta}^t$. Each visible satellite $k\in \mathcal{K}^t$ performs $E$ steps of local training (e.g., stochastic gradient descent or SGD) and produces either a local model $\boldsymbol{\theta}_{v,k}^t$ or its gradient $\mathbf{g}_{v,k}^t$. For example, after one local epoch,
\begin{equation}
\Resize{8.05 cm}{\mathbf{g}_{v,k}^t=\frac{1}{|\mathcal{D}_{v,k}|}
\sum_{(x,y)\in\mathcal{D}_{v,k}}\nabla f(\boldsymbol{\theta}^t;x,y),
\quad
\boldsymbol{\theta}_{v,k}^t = \boldsymbol{\theta}^t - \eta_t \mathbf{g}_{v,k}^t,}
\end{equation}
where $\eta_t$ is the local learning rate.

Each satellite then \emph{securely transmits} its update to one or more visible HAPs $h \in \mathcal{H}$, subject to link-capacity constraints (Sec.~\ref{sec:sys_mdl_visibility}). Before transmission, the update is protected using secure-aggregation primitives—e.g., HE, FE as in our prior work \cite{elmahallawy2023secure}, or MPC—yielding a ciphertext $\mathcal{C}_{v,k}^t$ along with a vendor-specific digital signature $\sigma_{v,k}^t$. 

Each HAP validates and records the received update on the blockchain ledger {(see Sec.~\ref{sec:method})} to provide a tamper-evident, auditable history of contributions. Once sufficient updates are collected, the HAPs exchange partial aggregates via inter-HAP links and compute a new global model $\boldsymbol{\theta}^{t+1}$. This process is repeated until a termination criterion is satisfied (e.g., convergence below a target loss or reaching a maximum number of rounds). Finally, the converged global model $\boldsymbol{\theta}^*$ is broadcast to all GSs so each vendor can deploy it within its constellation.

\subsection{Visibility and communication model} \label{sec:sys_mdl_visibility}
Satellite \(k\) is said to be \emph{visible} to HAP \(h\) at round \(t\) if there exists an active communication window between them (i.e., line-of-sight and sufficient link budget). We capture visibility with the binary indicator \(\gamma_{k,h}^t\in\{0,1\}\), where \(\gamma_{k,h}^t=1\) denotes visibility. A satellite may be visible to multiple HAPs simultaneously; the set of satellites visible to HAP \(h\) at round \(t\) is therefore $\mathcal{K}_h^t \;=\; \{\,k\in\mathcal{K}\;|\;\gamma_{k,h}^t=1\,\}$. Practical visibility requires meeting link constraints such as bandwidth \(B_{k,h}^t\) and pass duration \(\tau_{k,h}^t\); accordingly the transmitted payload must satisfy the capacity constraint $\mathrm{size} (\mathcal{C}_{v,k}^t)\;\le\; B_{k,h}^t\cdot\tau_{k,h}^t$, and may additionally be conditioned on minimum signal-to-noise ratio (SNR) or other quality-of-link thresholds as needed.

\subsection{Threat Model}\label{sec:threat_model}
We consider a multi-vendor LEO satellite network where each vendor operates its own satellites, HAPs, and GSs. The system assumes {\em honest-but-curious} and potentially malicious entities, including:  

\begin{itemize}[leftmargin=*]

\item {\bf \em Colluding vendor / compromised infrastructure.} A vendor-controlled satellite or a compromised shared $\mathcal {PS}$ (i.e., HAP) that can falsify metadata, over-weight contributions, censor or equivocate on commits, collude with other parties to bias aggregation, or exfiltrate confidential updates.
    
\item {\bf \em Active network attacker (MitM).} Intercepts, tampers, replays, or drops messages to disrupt aggregation, cause stale updates, or mount rollback attacks.
\item \textbf{Untrusted HAPs or GSs:} HAPs or GSs may attempt to tamper with, drop, or reorder model updates.
\item \textbf{Eavesdroppers:} Adversaries may attempt to intercept communication between satellite–HAP, HAP–GS, or inter-HAP links to extract sensitive data.
\item {\bf \em Off-chain storage adversary.} Compromises off-chain repositories holding bulk encrypted models or artifacts to tamper with or attempt to access model contents.
\end{itemize}
 \textbf{\em Adversary capabilities.} We denote an adversary $\mathcal{A}$ with corruption budget $(f_V,f_H,f_K)$ meaning up to $f_V$ vendors, $f_H$ HAPs, and $f_K$ satellites may be corrupted; $\mathcal{A}$ may collude and control some network links but is computationally bounded under standard cryptographic assumptions.

 \subsection{Security Goals}
\noindent \textsc{OrbitChain} is designed to meet the following security objectives, assuming corruption budgets respect consensus and cryptographic thresholds (e.g., $f_H < H/3$ for BFT consensus and appropriate $t$-of-$n$ for threshold cryptography):

\begin{enumerate}[label= {\bf \em (O\arabic*)}, leftmargin=0pt, labelwidth=*, labelsep=0.5em, align=left]
\item \textbf{Integrity \& provenance:} Every model update and its metadata are attributable to a specific participant and recorded immutably to prevent retroactive tampering.
  \item \textbf{Confidentiality:} Raw sensor data and vendor-sensitive information cannot be recovered from exchanged updates or off-chain artifacts.
  \item \textbf{Availability:} Aggregation and model distribution remain timely and resistant to censorship or denial-of-service on critical links and HAPs.
  \item \textbf{Auditability \& Accountability:} Tamper-evident logs enable post-hoc forensics, dispute resolution, and punishment of misbehavior (e.g., slashing or reputation updates).
    \item \textbf{Immutability:} Once a commit or aggregation event is finalized on-chain, it is computationally infeasible for any coalition below the validator threshold to alter or erase it.

\end{enumerate}

 \begin{figure*}[!t]
  \centering
  \includegraphics[width=1\linewidth]{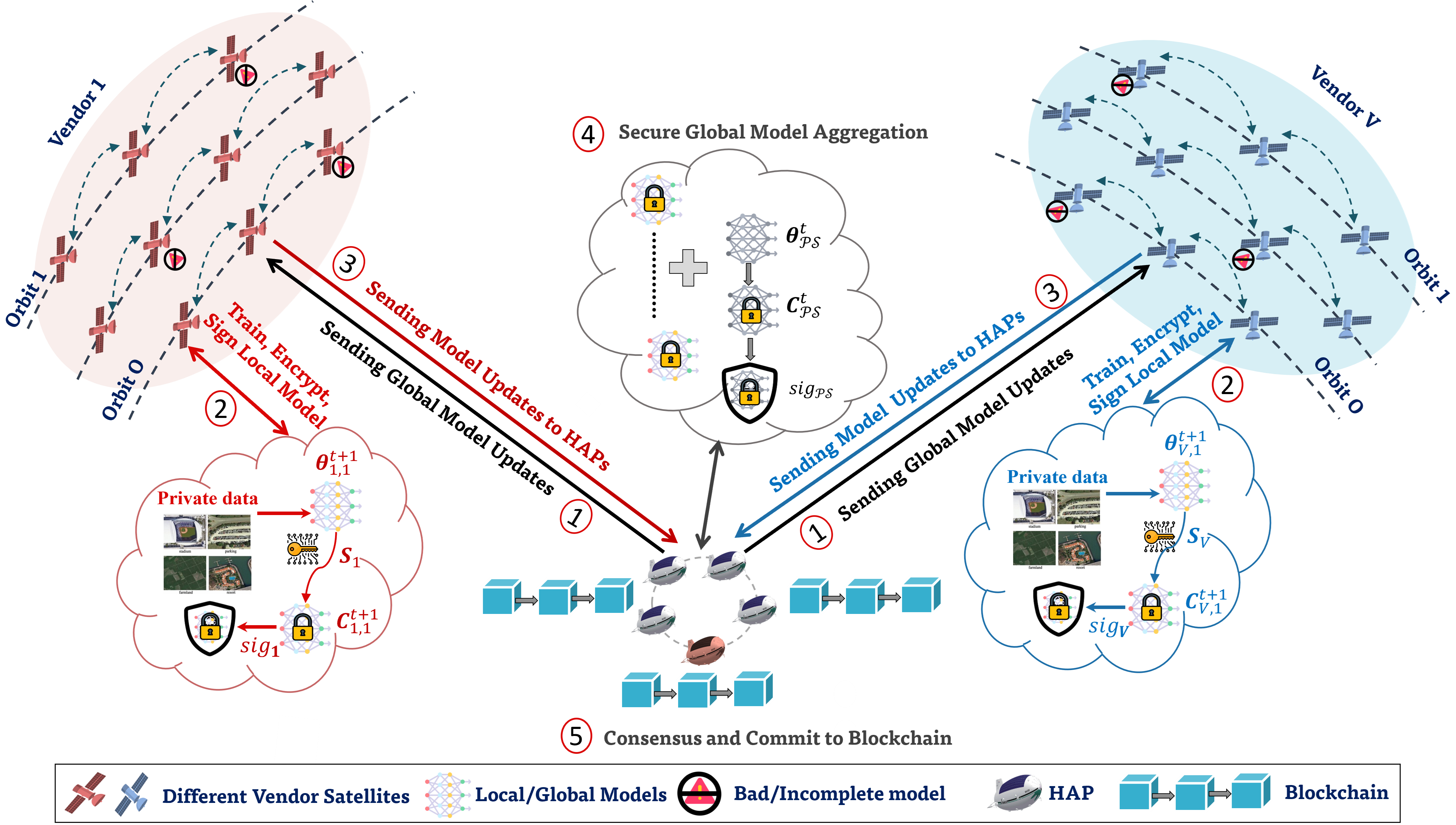}
  \caption{An illustration of the \textsc{OrbitChain} Framework.}
  \label{fig:framework}
\end{figure*}

\section{Proposed \textsc{OrbitChain} Framework}\label{sec:method}

This section presents \textsc{OrbitChain}, a technically grounded framework that enables privacy-preserving, auditable, and incentive-aligned multi-vendor FSL. \textsc{OrbitChain} is composed of two tightly integrated components: (i) a \emph{Cross-Vendor Model Optimization Layer}, which coordinates hierarchical communication across satellites, HAPs, and inter-HAP links within and across vendors. This layer fuses vendor-specific global models into a unified, generalized global model, thereby improving convergence efficiency by reducing the number of FSL communication rounds typically required to reach comparable accuracy, while maintaining high model performance even on unseen classes through effective cross-vendor collaboration; and (ii) a \emph{Blockchain-Backed Consensus Layer}, which ensures provenance, supports commit–reveal operations, and provides lightweight on-chain governance. This layer empowers satellites from different vendors—each operating their own {\em terrestrial} (i.e., GSs) and {\em non-terrestrial} (i.e., HAPs) infrastructure—to contribute model updates in a verifiable, tamper-evident, and transparent manner. The following subsections describe the key mechanisms and design principles of \textsc{OrbitChain} in concise technical form. 




\subsection{Cross-Vendor Model Optimization Layer} \label{sec:methodlogy}
This \textsc{OrbitChain} layer enables a hierarchical, efficient, and fused aggregation pipeline across satellites belonging to multiple vendors. It (i) accelerates convergence on heterogeneous cross-vendor data by enabling coordinated model fusion through HAP-level collaboration, where HAPs act as in-space aggregators that leverage superior satellite visibility and reduced round-trip latency to efficiently synchronize updates across vendors; and (ii) preserves vendor privacy by confining raw data and intermediate updates within each vendor’s own satellite and $\mathcal{PS}$ domain, while employing age- and reputation-aware per-satellite weighting to ensure fair contribution and mitigate the impact of stale or low-quality updates. The process proceeds through several stages as follows (see \fref{fig:framework}).

\begin{itemize}[leftmargin=*]

\item \textbf{Stage 1: Global Model Distribution.} At the beginning of round $t$, each HAP $h \in \mathcal{H}$ holds a synchronized copy of the current global model $\boldsymbol{\theta}^t$. 
Once a satellite $k$ (belonging to vendor $v$) becomes visible to a HAP $h$, it downloads this model:
$\boldsymbol{\theta}_{v,k}^{t,0} \leftarrow \boldsymbol{\theta}_h^t$. 


\vspace{0.5em}

\item\textbf{Stage 2: Local Model Training.} Each visible satellite $k$ updates its model locally using its private dataset $\mathcal{D}_{v,k}$ for $E$ epochs to compute model updates:
\begin{align}
    \boldsymbol{g}_{v,k}^t &= \frac{1}{|\mathcal{D}_{v,k}|} \sum_{(x,y)\in\mathcal{D}_{v,k}} \nabla f(\boldsymbol{\theta}_{v,k}^{t,e}; x, y),\\ \nonumber
    \boldsymbol{\theta}_{v,k}^{t+1,e+1} &= \boldsymbol{\theta}_{v,k}^{t,e} - \eta_t \boldsymbol{g}_{v,k}^t, \qquad e=0,\dots,E-1
\end{align}
  
To protect model parameters, each satellite encrypts its updated model $\boldsymbol{\theta}_{v,k}^{t+1}$ using an encryption scheme for secure aggregation (e.g., FE \cite{elmahallawy2023secure}):
\begin{equation}
    \mathcal{C}_{v,k}^{t+1} = \textsf{Enc}_{v}(\boldsymbol{\theta}_{v,k}^{t+1}),
\end{equation}
The ciphertext $\mathcal{C}_{v,k}^{t+1}$ is accompanied by a vendor signature $\sigma_{v,k}^{t+1} = \textsf{Sign}_v(\mathcal{C}_{v,k}^{t+1})$ and a compact metadata record containing the data size $|\mathcal{D}_{v,k}|$, timestamp, and vendor identifier. This metadata is committed on-chain as a cryptographic digest, serving as a vendor-signed Proof-of-Authority (PoA) and ensuring verifiable provenance of each model update, detailed in \sref{sec:blkchn}.

\vspace{0.5em}

\item\textbf{Stage 3: Satellite-HAP-Transmission and Commit.} Each visible satellite $k_h^t \in \mathcal{K}_h^t$ in round $t$ then transmits its encrypted update $\mathcal{C}_{v,k}^{t+1}$ to the corresponding HAP $h$, subject to the link-capacity constraint:
\begin{equation}
\mathrm{size}(\mathcal{C}_{v,k}^{t+1}) \le B_{k,h}^t \, \tau_{k,h}^t,
\end{equation}
which accounts for the available bandwidth $B_{k,h}^t$ and the communication window duration $\tau_{k,h}^t$. Upon successful reception, the HAP $h$ computes a compact digest:
\begin{equation}
    d_{v,k}^{t} \;=\; \mathcal{H}\big(\mathcal{C}_{v,k}^{t+1} \,\|\, \sigma_{v,k}^{t+1} \,\|\, \text{meta}_{v,k}^{t+1}\big),
\end{equation}
and proposes $d_{v,k}^t$ for inclusion in the next block (\sref{sec:blkchn}). 

\item\textbf{Stage 4: HAP-Level Secure Aggregation.} After commit and (if required) a controlled reveal via threshold-decryption shares, the HAP $h$ computes a \emph{secure local aggregate} from its visible satellites $\mathcal{K}_h^t$ as:
\begin{equation}\label{eq:hap_local_agg}
    \boldsymbol{\theta}_h^{t+1} = \sum_{k \in \mathcal{K}_h^t} \alpha_{v(k),k}^t \, Q(\textsf{Dec}_h(\mathcal{C}_{v,k}^t)),
\end{equation}
where $Q(\cdot)$ optionally applies compression or quantization for bandwidth efficiency. If full decryption is restricted for privacy, partial decryption shares are combined via threshold cryptography or MPC among collaborating HAPs. $\alpha_{v,k}^t$ denotes the \emph{normalized, age- and reputation-aware per-satellite weight}, designed to ensure fairness and robustness, and is defined as:
\begin{equation}
\alpha_{v,k}^t \;=\; 
\frac{ |\mathcal{D}_{v,k}| \cdot r_{v,k}^t \cdot \mathrm{decay}\!\big(a_{v,k}^t\big) }
{\sum\limits_{k'\in\mathcal{K}_h^t} |\mathcal{D}_{v(k'),k'}| \cdot r_{v(k'),k'}^t \cdot \mathrm{decay}\!\big(a_{v(k'),k'}^t\big)},
\end{equation}
where $r_{v,k}^t$ represents the \emph{reputation score} of satellite $k$ belonging to vendor $v$, reflecting its historical reliability and contribution quality, and $\mathrm{decay}(a)$ models temporal freshness (representing update staleness), typically as an exponential function $\mathrm{decay}(a)=\exp(-\lambda a)$.

\textbf{\em Remarks:} Eqn. \eqref{eq:hap_local_agg} allows each HAP to (i) discount stale updates, (ii) prefer higher-reputation contributors, and (iii) normalize by local data volume to reduce bias from vendors with disproportionately large constellations.

\vspace{0.5em}

\item\textbf{Stage 5: Cross-HAP Consensus and Global Fusion.}
After local aggregation, HAPs exchange their local aggregates (or their digests) and invoke the ledger's consensus to (i) commit the set of local aggregates and (ii) agree on a reconciliation policy. Then, all HAPs engage in ledger-backed consensus (see \sref{sec:blkchn}) to fuse their local aggregates into the global model:
\begin{equation}
\boldsymbol{\theta}^{t+1} = \sum_{h \in \mathcal{H}} \beta_h^t \, \boldsymbol{\theta}_h^{t+1},
\end{equation}
 where $\beta_h^t$ is chosen to reflect the total participating data behind HAP $h$, the freshness of its updates (inverse age), and per-vendor fairness constraints, which can be given as: 
\begin{equation}
\beta_h^t \;=\; \frac{\sum_{k\in\mathcal{K}_h^t} |\mathcal{D}_{v(k),k}|\cdot \mathrm{decay}(a_{v(k),k}^t)}{\sum_{h'} \sum_{k\in\mathcal{K}_{h'}^t} |\mathcal{D}_{v(k),k}|\cdot \mathrm{decay}(a_{v(k),k}^t)}.
\end{equation}
Each aggregation result is recorded on the blockchain to ensure \emph{immutability} and \emph{cross-vendor auditability}.

Once $\boldsymbol{\theta}^{t+1}$ is finalized by the HAP committee and committed on-chain (e.g., aggregate digest and contributor list), each HAP pushes $\boldsymbol{\theta}^{t+1}$ to its visible satellites or multicasts it to its GSs for distribution. All digests, contribution metadata, and reveal events remain immutably logged for auditability.

\end{itemize}

\subsection{\textsc{OrbitChain}'s Blockchain-Based Layer}\label{sec:blkchn}

This subsection presents the blockchain component of \textsc{OrbitChain}, which employs a \emph{flexibly coupled blockchain–federated learning} (FlC-BCFL) design. The blockchain layer functions as an independent provenance and aggregation substrate—separate from model training—to ensure transparency, traceability, and tamper-evident operation. Unlike \emph{fully coupled} frameworks that require every satellite to act as a validator, \textsc{OrbitChain} delegates consensus exclusively to high-altitude platforms (HAPs), which serve as computationally capable, energy-stable, and consistently connected validators. This separation allows satellites to remain lightweight while still participating in a system that enforces verifiable accountability for every model update.

\textsc{OrbitChain} employs a permissioned blockchain governed by a Proof of Authority (PoA) consensus mechanism. In PoA, block validation is performed by a predetermined set of trusted authorities—here, the HAP validators—rather than through resource-intensive mechanisms such as Proof of Work (PoW) or stake-based economics as in Proof of Stake (PoS). Each HAP validator possesses a cryptographic identity, publicly registered in the \texttt{OrbitLedger} contract, and is responsible for signing and confirming new blocks. A block is finalized when a threshold number of validators, typically $f_H < H/3$ faulty nodes tolerated under Byzantine assumptions, co-sign the block. This guarantees deterministic finality and immediate confirmation, a property critical for low-latency federated learning rounds.

Mathematically, let $V_h$ represent the set of validators and $\mathcal{S}_h$ their signatures. A block proposal $B_t$ at round $t$ is finalized when:
\[
|\mathcal{S}_h| \geq \tau, \quad \text{where} \quad \tau = \left\lceil \frac{2H}{3} \right\rceil.
\]
This quorum-based rule ensures that the system remains live and consistent as long as fewer than one-third of validators are compromised, following the safety guarantees of Practical Byzantine Fault Tolerance (PBFT) but at significantly lower communication overhead.

PoA is particularly well-suited for \textsc{OrbitChain} for three reasons: (i) HAPs are semi-trusted entities that can be authenticated by their vendors through verifiable certificates; (ii) the deterministic finality of PoA eliminates the probabilistic delays common in PoW or PoS networks, aligning with the short time windows available in LEO communications; and (iii) PoA’s lower message complexity, $O(n)$ per block compared to PBFT’s $O(n^2)$, minimizes communication overhead in bandwidth-limited space networks.

\textsc{OrbitChain} integrates this PoA consensus with a lightweight ledger design that records only compact, verifiable artifacts—hashes, signatures, and metadata—on-chain, while storing full encrypted models off-chain. A single persistent smart contract, \texttt{OrbitLedger}, maintains public keys for vendors, satellites, and HAP validators and logs round-level evolution through indexed on-chain events.

Each contribution—local update, partial aggregate, or global model—is represented by a compact token:
\[
K = \mathcal{H}\big(\mathrm{vendorID}\,\|\,\mathrm{satID}\,\|\,t\,\|\,\tau\,\|\,\mathcal{H}(X)\big),
\]
binding it cryptographically to its source, timestamp, and content hash. Two rolling Merkle accumulators summarize blockchain state: $M_{\text{round}}[t]$ commits artifacts for the current round, and $M_{\text{vendor}}[v]$ aggregates contributions by vendor. These enable logarithmic-time verification without full chain traversal.

During each round, encrypted updates $\mathcal{C}_{v,k}^t$ and their metadata are transmitted from satellites to their visible HAPs. Upon validation, each HAP computes a digest:
\[
d_{v,k}^{t} = \mathcal{H}\big(\mathcal{C}_{v,k}^t \,\|\, \sigma_{v,k}^t \,\|\, \mathrm{meta}_{v,k}^t\big),
\]
emits a \texttt{Commit} event, and co-signs the block proposal. Once the quorum $\tau$ is reached, the block is finalized, ensuring that all commits within that round are immutable. Subsequent \texttt{PartialAgg} and \texttt{GlobalAgg} events finalize aggregation, while a \texttt{Distribute} event records the off-chain model’s storage location for auditability and later verification.

Finally, provenance verification occurs entirely through the blockchain’s event log: auditors can trace every global model token $K_{\mathrm{glob}}$ back through its hierarchical lineage to the individual vendor updates that formed it. This lineage not only supports post-hoc forensics but also guarantees that each committed update (i) was registered before aggregation, (ii) originated from an authenticated vendor, and (iii) contributed exactly once to the final model.

Together, the PoA consensus and \texttt{OrbitLedger} architecture create a deterministic, low-latency, and verifiably fair environment for federated learning across multi-vendor satellite networks—ensuring integrity without the computational burden of traditional consensus mechanisms.

\section{Performance Evaluation}

\subsection{Experimental Setup}

\noindent{\bf Muli-vendor Network.} We simulate a realistic multi-vendor LEO FSL environment that can be reproduced with the \texttt{skyfield} Python library. The scenario parameters are:

\begin{itemize}[leftmargin=*]

\item \textbf{Vendors and constellations:} We model three vendors (Kuiper,  OneWeb, Starlink) using simplified constellations of \(K_v = 21\) satellites over \(P = 3\) orbital planes (7 per plane).  Planes are evenly spaced in RAAN, and satellites within each plane are uniformly phased.

  
  \item \textbf{Orbit geometry:} Nominal circular altitude $h=550\,$km (typical LEO); vendor-specific inclinations to generate heterogeneity, e.g., $i_1=53^\circ$, $i_2=97.5^\circ$, $i_3=30^\circ$. 
  These settings are purely for simulation and do \emph{not} reflect real vendors’ orbital inclinations, altitudes, or constellation sizes.

  \item \textbf{HAPs and GSs:} Each vendor has three HAPs and one GS. HAPs are fixed at stratospheric altitudes (e.g., 20\,km) with chosen lat/lon coordinates. All HAPs (across vendors) form a full inter-HAP mesh; GSs are vendor-owned and \emph{not} directly interconnected across vendors.
  
  \item \textbf{Communications / link model:} A satellite $k$ is visible to HAP $h$ at time $t$ if elevation$(k,h,t) \ge \theta_{\text{min}}$ (use $\theta_{\text{min}}=10^\circ$). For each satellite–HAP pair, we record pass duration $\tau_{k,h}^t$ and available bandwidth $B_{k,h}^t$, and enforce $\mathrm{size}(\mathcal{C}_{v,k}^t) \le B_{k,h}^t \cdot \tau_{k,h}^t$. Use a per-pass bandwidth (e.g., $B=10\,$Mbps) as a starting point and sample sensitivity in experiments.
\end{itemize}

\vspace{1mm}
\noindent\textbf{Training Settings \& Datasets.} We evaluate \textsc{OrbitChain} under two training regimes: an \emph{i.i.d.} setting in which each satellite receives an equal share of the global dataset with all classes represented, and a \emph{non-i.i.d.} (label-skew) setting in which each satellite is assigned data from only two classes. The non-i.i.d. split emulates highly heterogeneous sensing aboard vendor constellations and stresses cross-vendor generalization. 

We use three benchmark datasets, including real-world satellite imagery, and train models appropriate to each task:

\begin{itemize}[leftmargin=*]
  \item \textbf{MNIST}~\cite{deng2012mnist}: $70{,}000$ grayscale images of handwritten digits (10 classes). Satellites train a small two-layer convolutional neural network (CNN) for fast on-device learning and comparison with common FSL baselines.
  \item \textbf{EuroSat}~\cite{helber2019eurosat}: $\sim\!27{,}000$ RGB and multispectral satellite images spanning 10 land-use classes. We use the same two-layer CNN as in MNIST to capture spatial patterns.

  \item \textbf{UC Merced}~\cite{yang2010bag}: $2{,}100$ high-resolution aerial scene images across 21 categories, offering high visual variability. Satellites train a deeper model (ResNet-18) to handle the increased scene complexity.
\end{itemize}
These datasets span simple, medium, and high-complexity classification tasks, allowing us to measure per-vendor accuracy, convergence speed, and robustness of \textsc{OrbitChain} under both i.i.d. and non-i.i.d. data partitions.

\noindent\textbf{Baselines.}  Since blockchain-enabled learning for LEO satellites remains nascent, we compare \textsc{OrbitChain} to the two most relevant blockchain---FL frameworks developed for satellite networks \cite{pokhrel2021blockchain,wu2024sharded}. In addition, to assess \textsc{OrbitChain}'s impact on convergence speed and cross-vendor performance, we evaluate two practical FSL settings: (i) single-vendor and (ii) multi-vendor without blockchain support. These baselines allow us to clearly demonstrate the benefits of enabling trustworthy multi-vendor collaboration, including improved accountability, coordinated model learning, and faster convergence across heterogeneous satellite vendors.

\noindent\textbf{Evaluation Metric.} We assess \textsc{OrbitChain}'s model integrity and validator behavior using metrics that capture PoA responsiveness, stability, and correctness within the integrated orbit–ledger simulation: i) \emph{Consensus latency}: Time to verify and finalize a block under different quorum settings (1-of-5, 3-of-5, 5-of-5), ii)
\emph{Latency stability}: Distribution and variance of finalization times across 1{,}000 blocks, iii) \emph{Quorum sensitivity}: Impact of increasing validator signatures on finalization time, iv) \emph{Fault tolerance}: Stability of majority quorums (e.g., 3-of-5) under slow or inconsistent validators, v)\emph{Ledger correctness}: Ensuring all validators append identical blocks in order, and vi) \emph{Integration feasibility}: Confirming PoA finalizes within satellite visibility windows. For \textsc{OrbitChain}'s convergence, we measure convergence speed and final global model accuracy.

\subsection{\textsc{OrbitChain}'s Blockchain Integration \& Validator Results} 

To realize \textsc{OrbitChain}'s permissioned ledger within the simulated LEO environment, we build a lightweight blockchain layer using Python’s \texttt{Flask} framework. Each HAP acts as an independent validator exposing REST endpoints for block proposal, validation, and synchronization, enabling a reproducible emulation of PoA consensus aligned with FSL rounds. Specifically, the \texttt{Flask} design provides a minimal HTTP-based interface for distributed coordination. Each validator maintains a local JSON ledger, a mempool of transactions from visible satellites, and cryptographic signing utilities via \texttt{hashlib} and \texttt{RSA}. When a block is proposed, the remaining HAPs verify its lineage, broadcast approvals, and append the block upon reaching the quorum threshold, mirroring permissioned chains such as Hyperledger Fabric but at substantially lower computational cost.

We employ a five-node validator committee to reflect a realistic inter-HAP mesh. The PoA quorum rule follows:
\[
q_{\textsf{commit}} > \frac{n+f}{2},
\]
with $n{=}5$, enabling three evaluation modes: (i) 1-of-5 (minimal latency), (ii) 3-of-5 (fault-tolerant with one Byzantine node), and (iii) 5-of-5 (maximal safety). These settings highlight the trade-off between approval thresholds and block finalization time.

Integrating this \texttt{Flask}-based ledger with the orbit-level visibility model yields a hybrid testbed combining real communication constraints with distributed consensus. Each on-chain event maps to a verifiable aggregation round, producing immutable provenance and demonstrating that PoA coordination is compatible with real-time satellite--HAP interactions.

\begin{figure}[!t]
    \centering
    \includegraphics[width=1\linewidth]{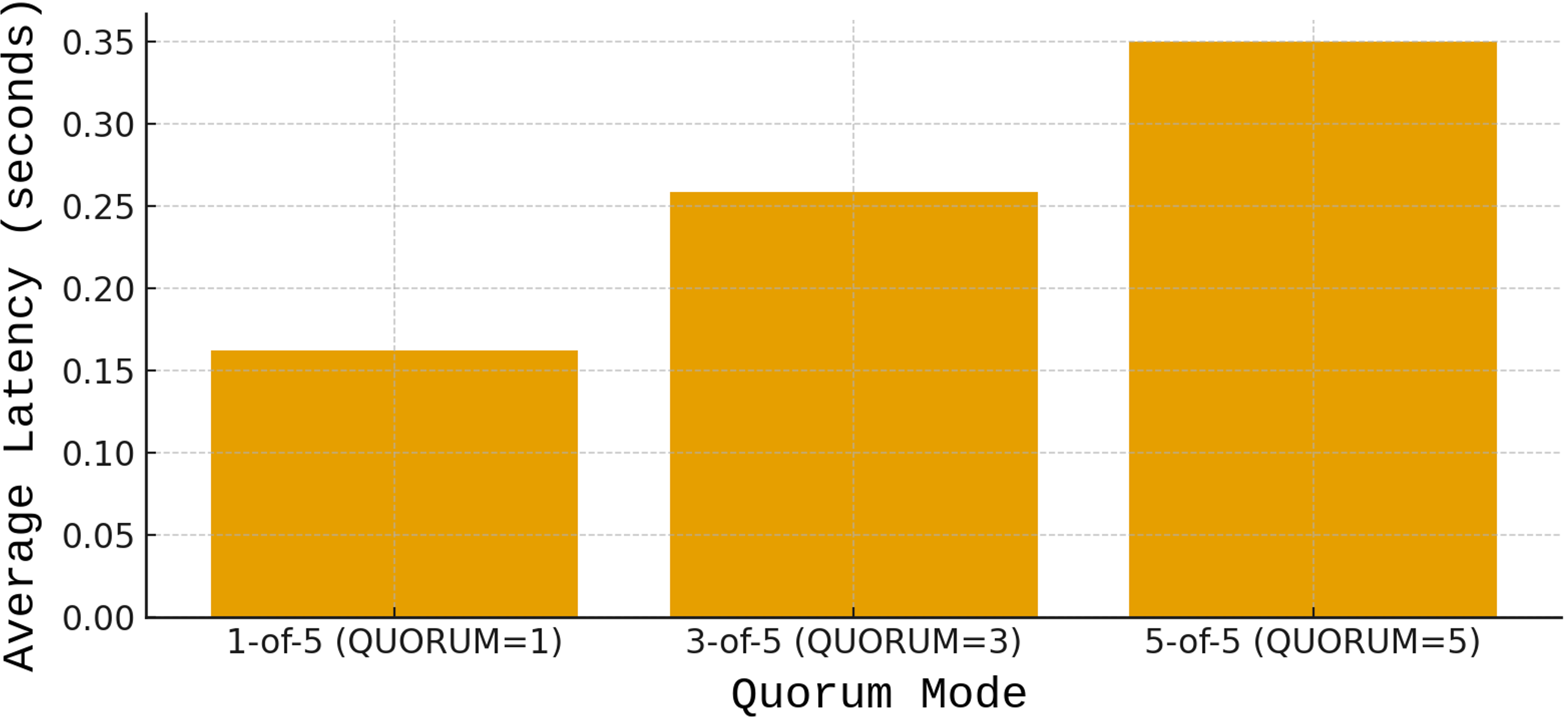}
    \caption{Average latency comparison across quorum modes (1-of-5, 3-of-5, 5-of-5).}
    \label{fig:avg_latency}
\end{figure}
\begin{figure}[!t]
    \centering
    \includegraphics[width=1\linewidth]{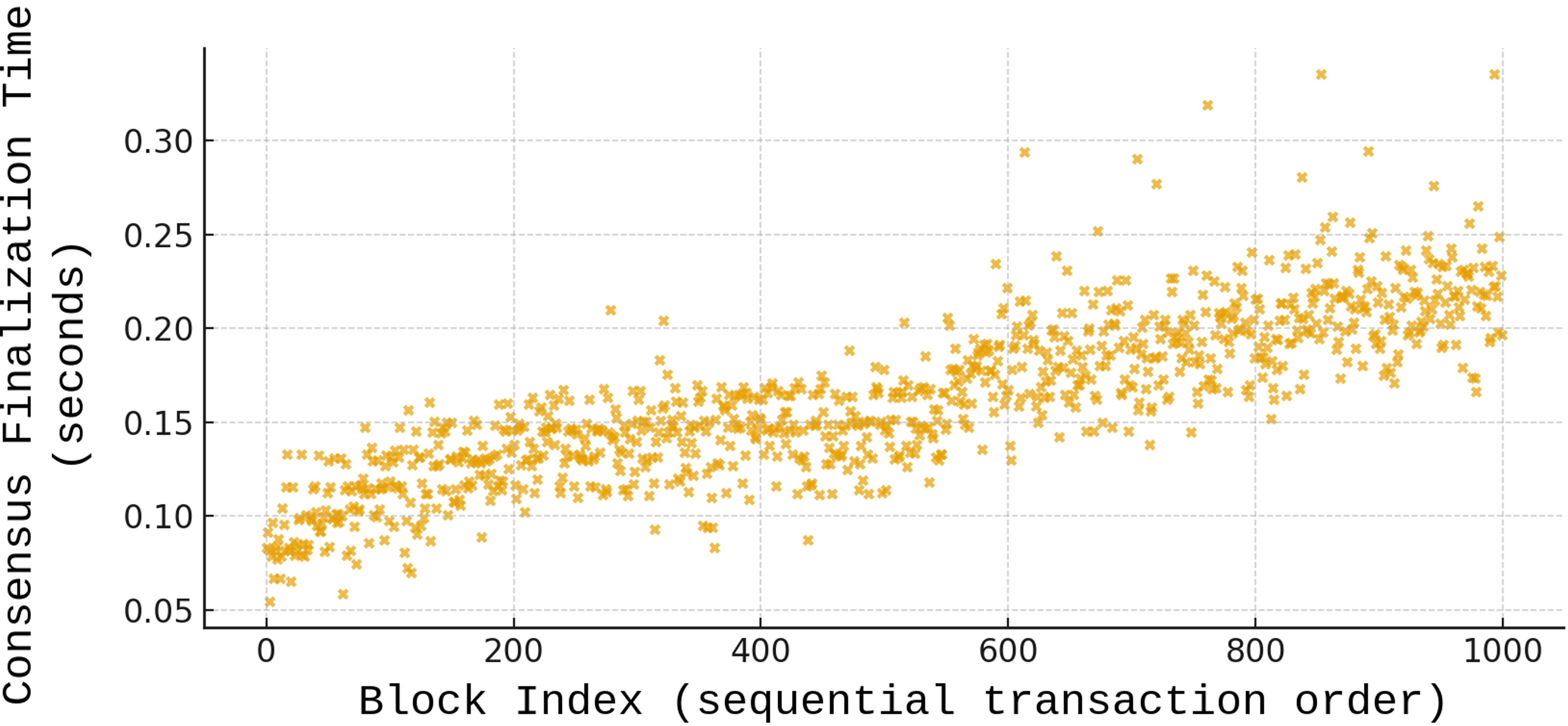}
    \caption{PoA performance under 1-of-5 quorum mode. Each point represents the consensus finalization time for a block.}
    \label{fig:poa1}
\end{figure}

\begin{figure}[!t]
    \centering
    \includegraphics[width=1\linewidth]{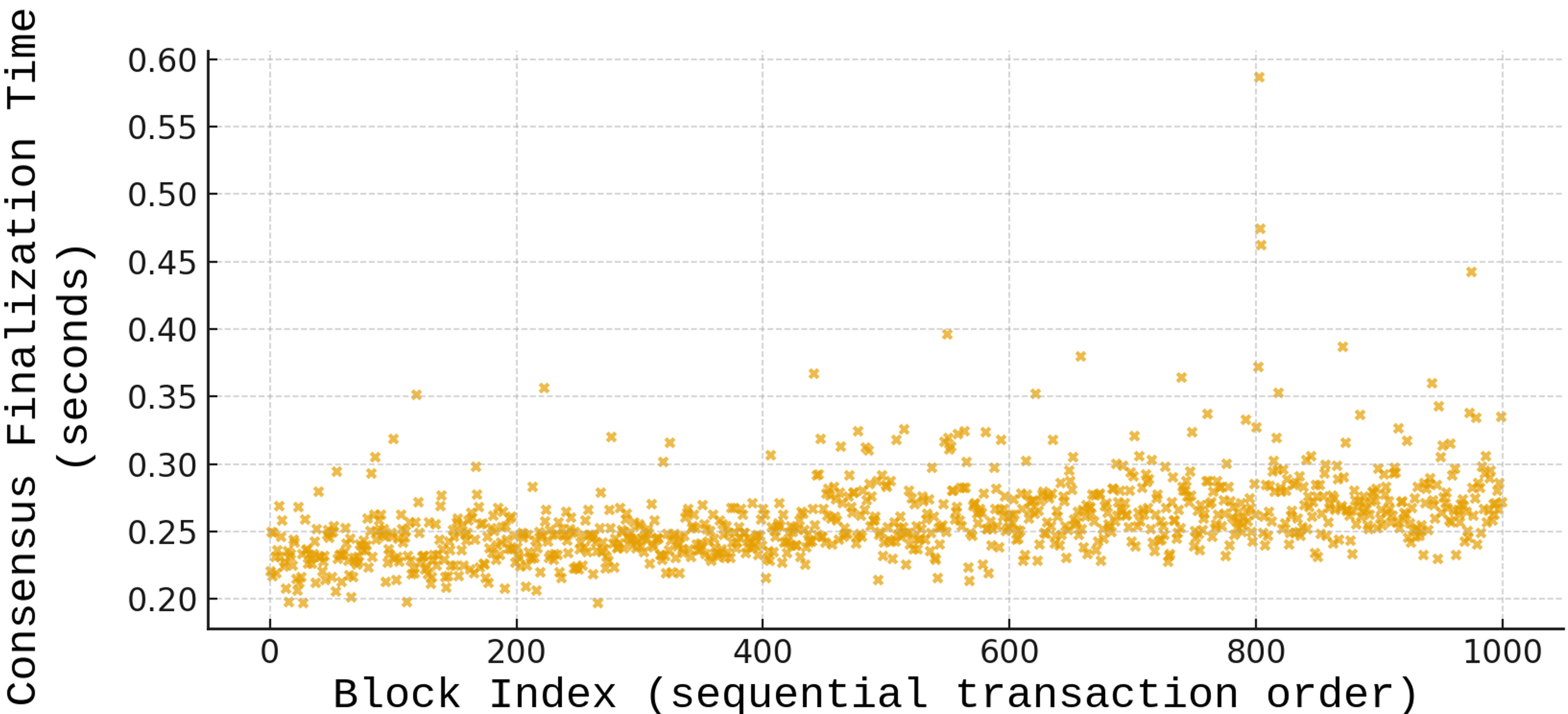}
    \caption{PoA performance under 3-of-5 quorum mode.}
    \label{fig:poa3}
\end{figure}
\begin{figure}[!t]
    \centering
    \includegraphics[width=1\linewidth]{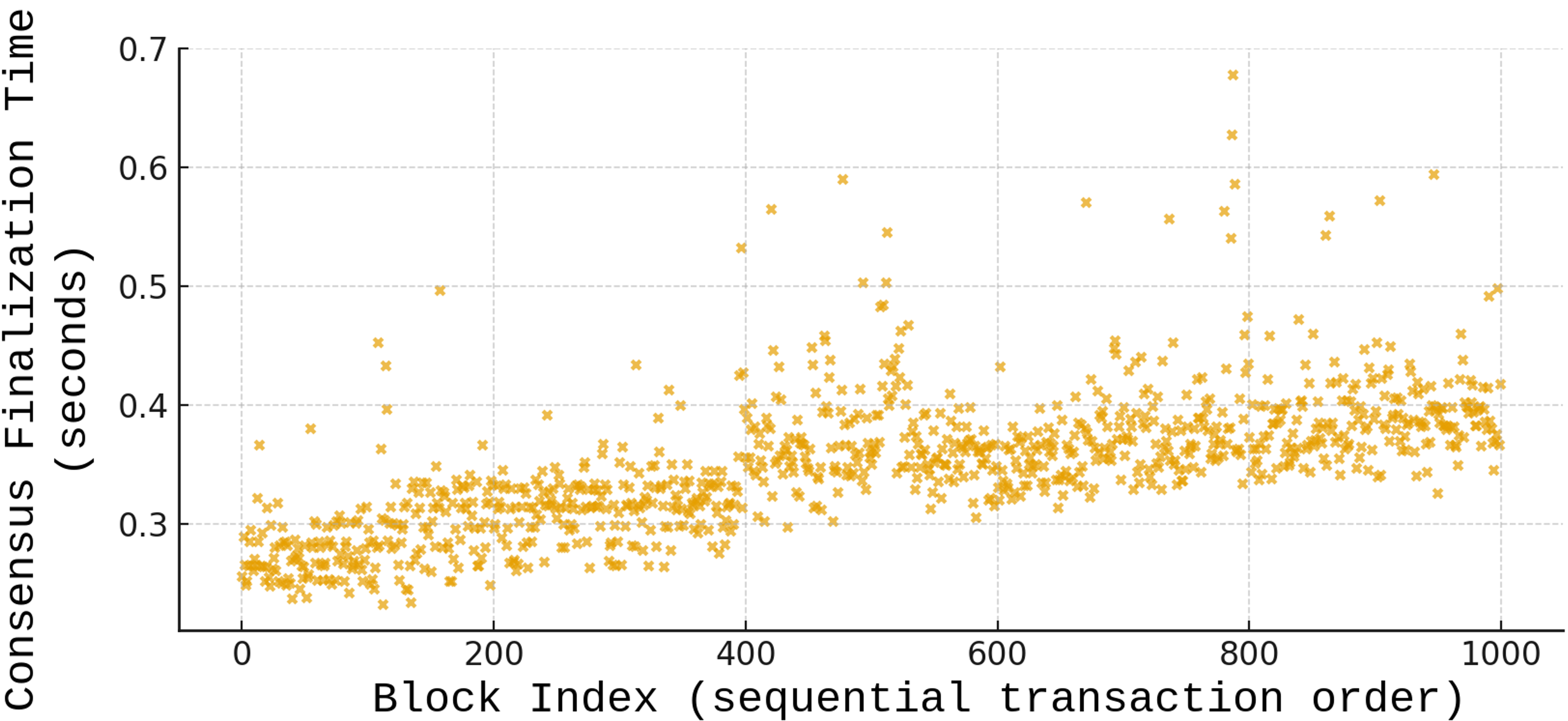}
    \caption{PoA performance under 5-of-5 quorum mode.}
    \label{fig:poa5}
\end{figure}

\begin{figure*}[!t]
  \centering
 
  \begin{subfigure}[b]{0.32\textwidth}
    \centering
    \includegraphics[width=\linewidth]{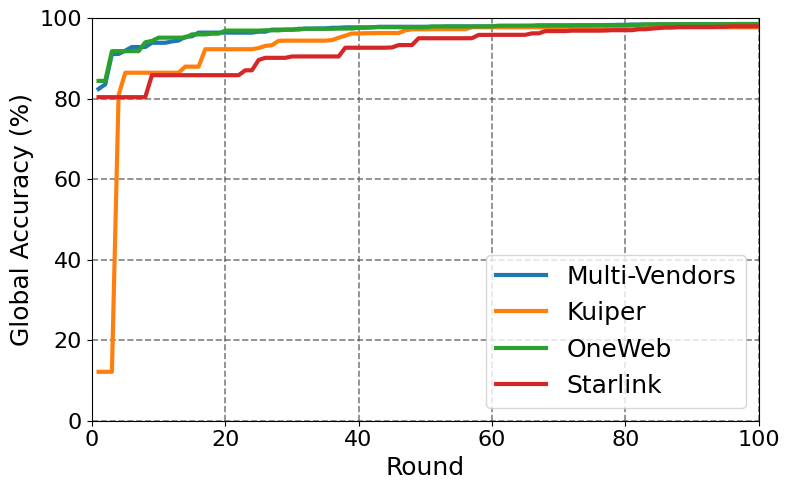}  
    \caption{10-minute slack time.}
  \end{subfigure}\hfill
  \begin{subfigure}[b]{0.32\textwidth}
    \centering
    \includegraphics[width=\linewidth]{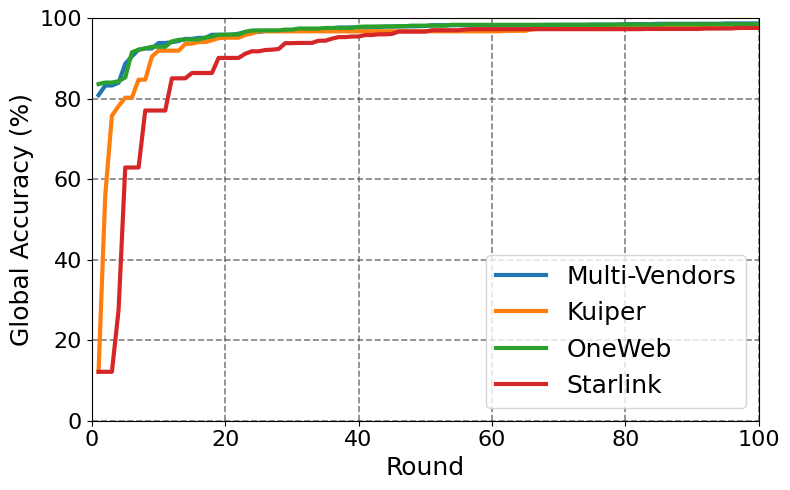} 
    \caption{20-minute slack time.}
  \end{subfigure}\hfill
  \begin{subfigure}[b]{0.32\textwidth}
    \centering
    \includegraphics[width=\linewidth]{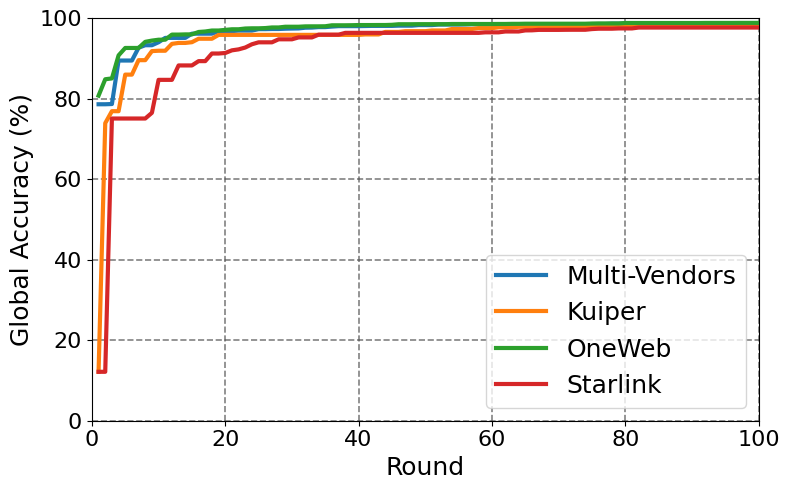} 
    \caption{30-minute slack time.}
  \end{subfigure}

  \begin{subfigure}[b]{0.32\textwidth}
    \centering
    \includegraphics[width=\linewidth]{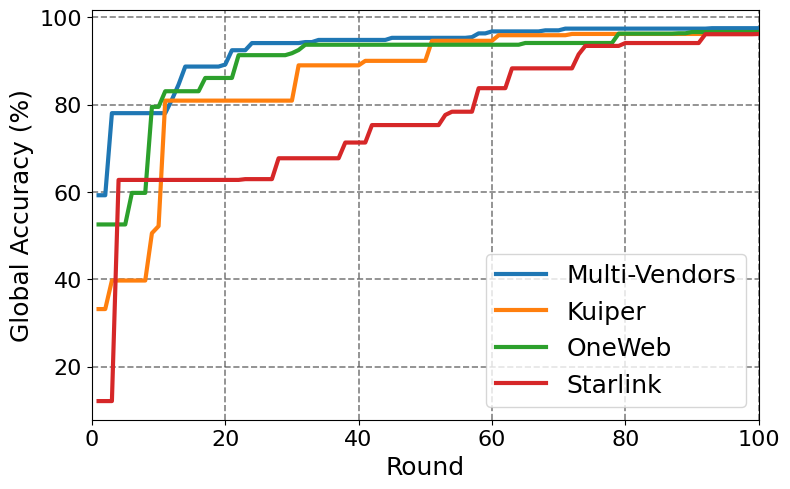} 
   \caption{10-minute slack time.}
  \end{subfigure}\hfill
  \begin{subfigure}[b]{0.32\textwidth}
    \centering
    \includegraphics[width=\linewidth]{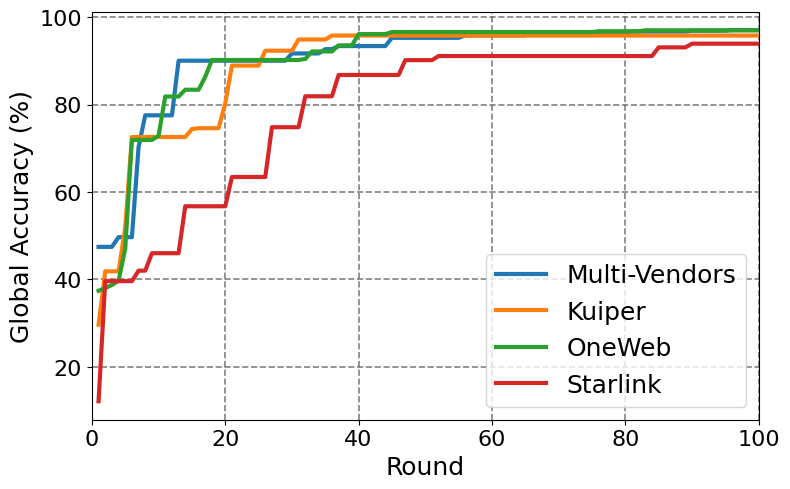} 
    \caption{20-minute slack time.}
  \end{subfigure}\hfill
  \begin{subfigure}[b]{0.32\textwidth}
    \centering
    \includegraphics[width=\linewidth]{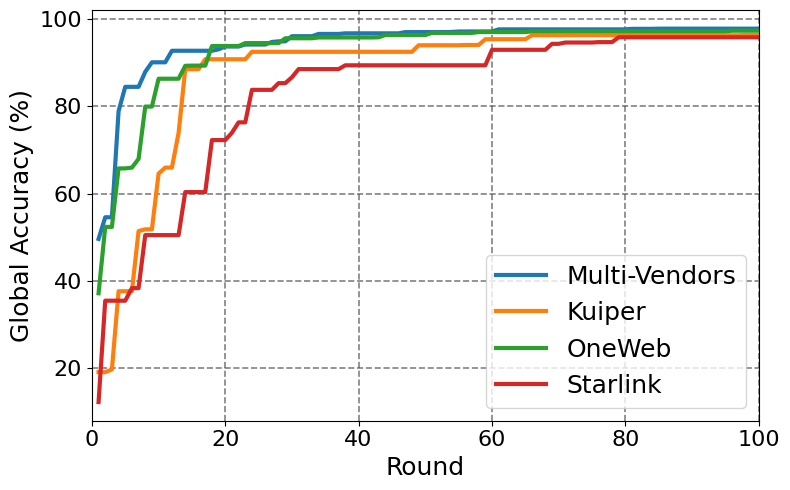} 
    \caption{30-minute slack time.}
  \end{subfigure}

\caption{\textsc{OrbitChain} convergence speed on the \textbf{MNIST} dataset under different slack times (10, 20, and 30 minutes) and varying data heterogeneity levels. Subfigures (a)--(b) present results for the i.i.d.\ setting, while subfigures (c)--(d) show the corresponding non-i.i.d.\ performance.}

  \label{fig:MNIST_Conv}
\end{figure*}

\begin{figure*}[!t]
  \centering
 
  \begin{subfigure}[b]{0.32\textwidth}
    \centering
    \includegraphics[width=\linewidth]{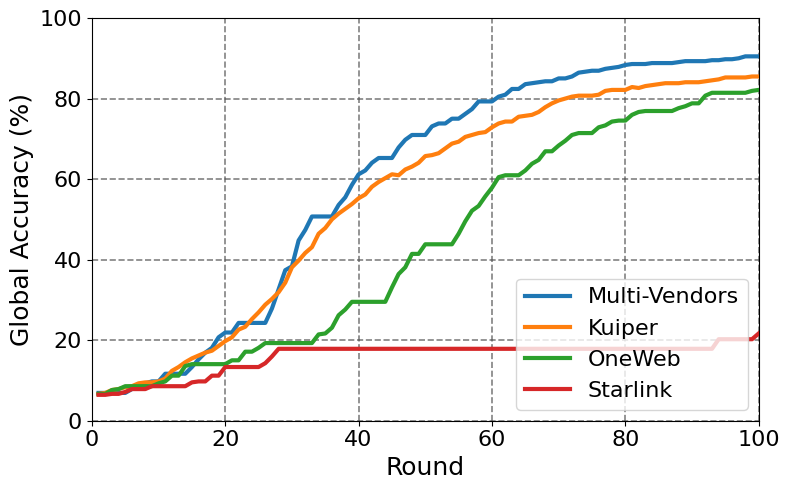} 
  \caption{10-minute slack time.}
  \end{subfigure}\hfill
  \begin{subfigure}[b]{0.32\textwidth}
    \centering
    \includegraphics[width=\linewidth]{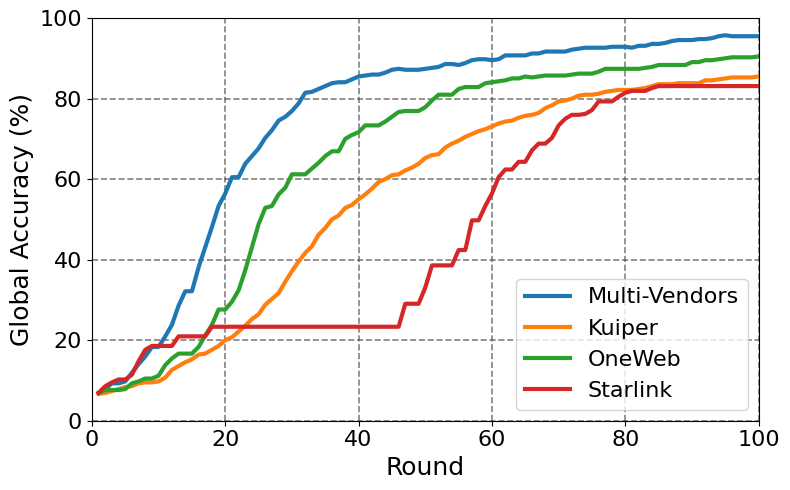} 
    \caption{20-minute slack time.}
  \end{subfigure}\hfill
  \begin{subfigure}[b]{0.32\textwidth}
    \centering
    \includegraphics[width=\linewidth]{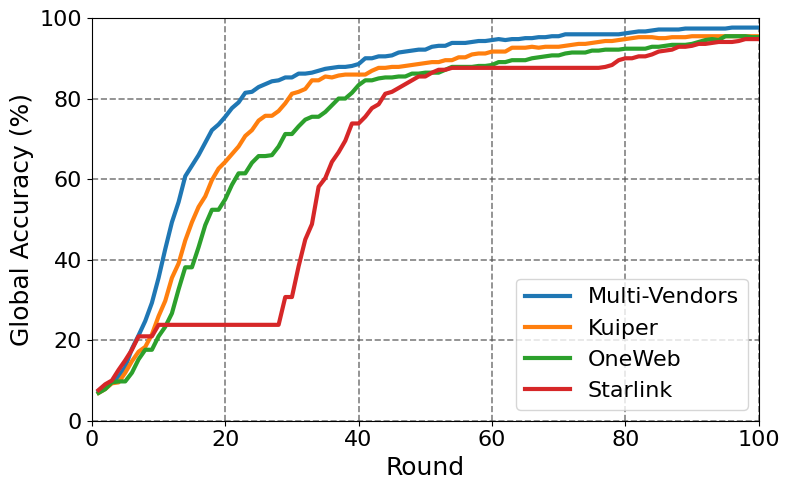} 
    \caption{30-minute slack time.}
  \end{subfigure}

  \begin{subfigure}[b]{0.32\textwidth}
    \centering
    \includegraphics[width=\linewidth]{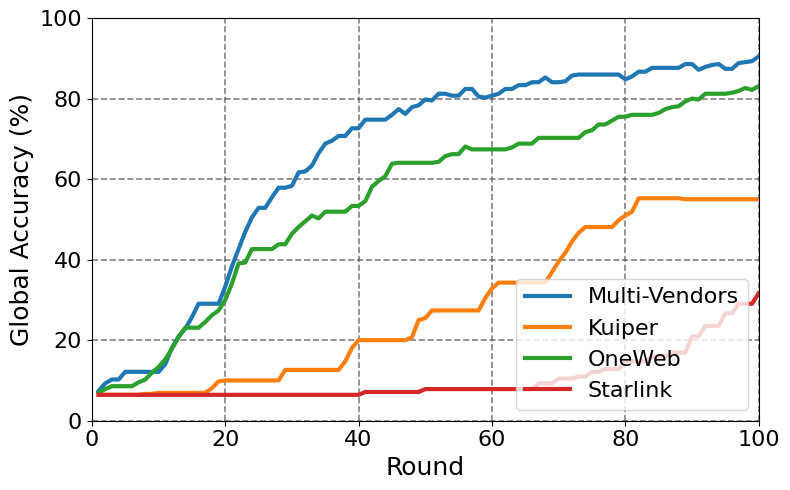}
    \caption{10-minute slack time.}
  \end{subfigure}\hfill
  \begin{subfigure}[b]{0.32\textwidth}
    \centering
    \includegraphics[width=\linewidth]{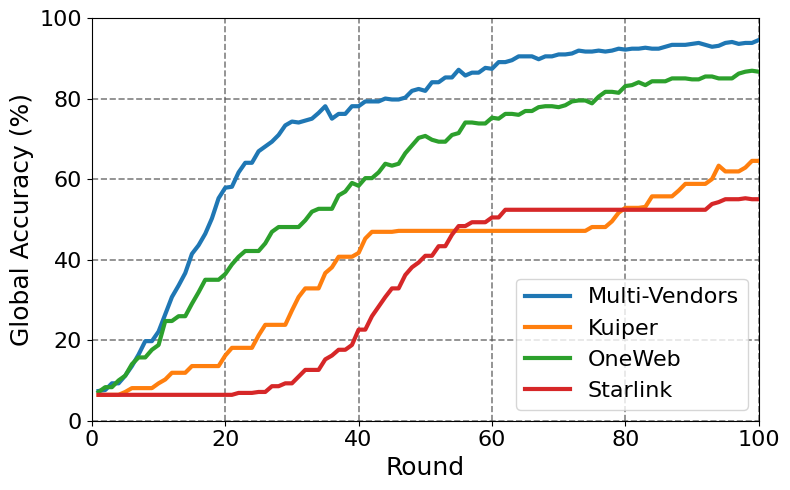} 
    \caption{20-minute slack time.}
  \end{subfigure}\hfill
  \begin{subfigure}[b]{0.32\textwidth}
    \centering
    \includegraphics[width=\linewidth]{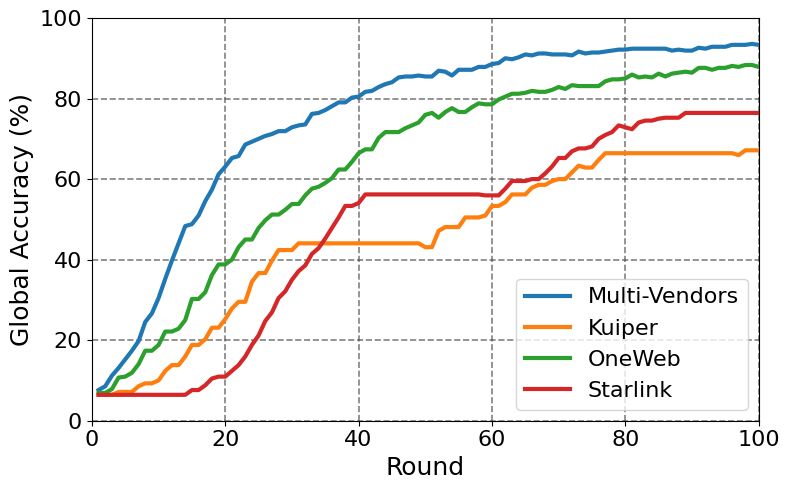} 
  \caption{30-minute slack time.}
  \end{subfigure}
\caption{\textsc{OrbitChain} convergence speed on the \textbf{UC Merced} dataset under different slack times (10, 20, and 30 minutes) and varying data heterogeneity levels. Subfigures (a)--(b) present results for the i.i.d.\ setting, while subfigures (c)--(d) show the corresponding non-i.i.d.\ performance.}
  \label{fig:UCMERCED_CONV}
\end{figure*}

\begin{figure*}[!t]
  \centering
  \begin{subfigure}[b]{0.32\textwidth}
    \centering
    \includegraphics[width=\linewidth]{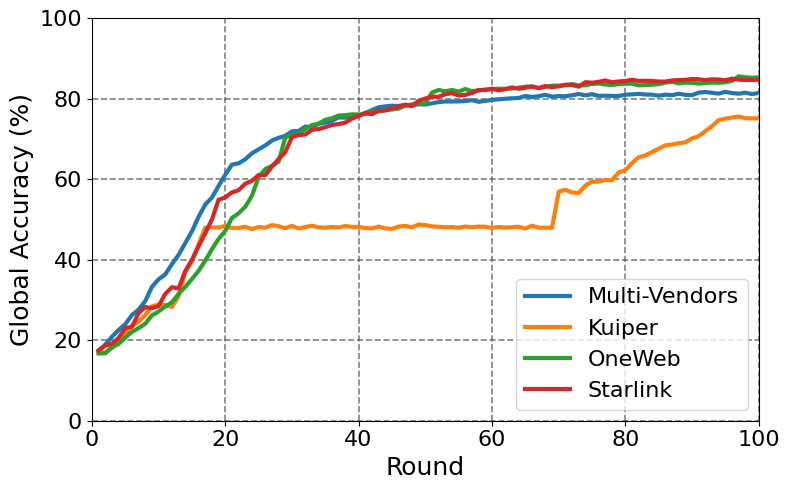} 
 \caption{10-minute slack time.}
  \end{subfigure}\hfill
  \begin{subfigure}[b]{0.32\textwidth}
    \centering
    \includegraphics[width=\linewidth]{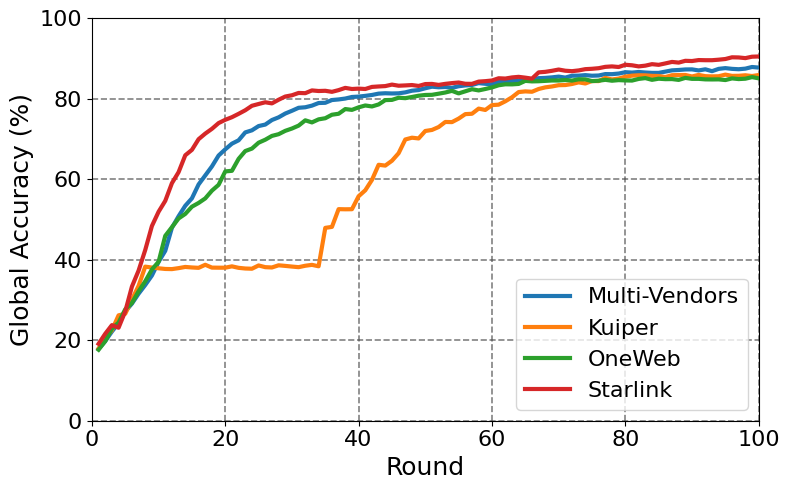} 
    \caption{20-minute slack time.}
  \end{subfigure}\hfill
  \begin{subfigure}[b]{0.32\textwidth}
    \centering
    \includegraphics[width=\linewidth]{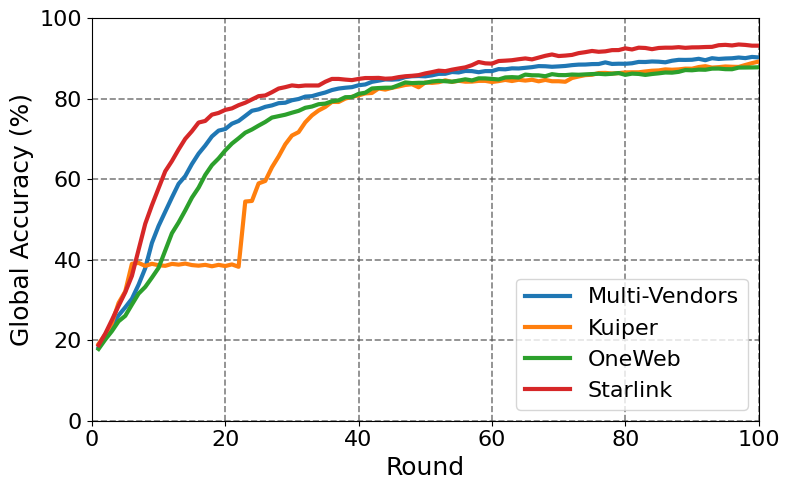} 
   \caption{30-minute slack time.}
  \end{subfigure}
  \begin{subfigure}[b]{0.32\textwidth}
    \centering
    \includegraphics[width=\linewidth]{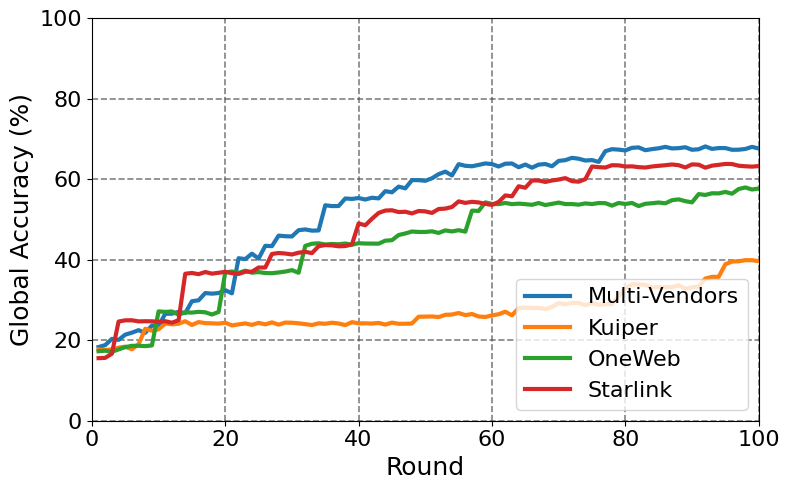}
    \caption{10-minute slack time.}
  \end{subfigure}\hfill
  \begin{subfigure}[b]{0.32\textwidth}
    \centering
    \includegraphics[width=\linewidth]{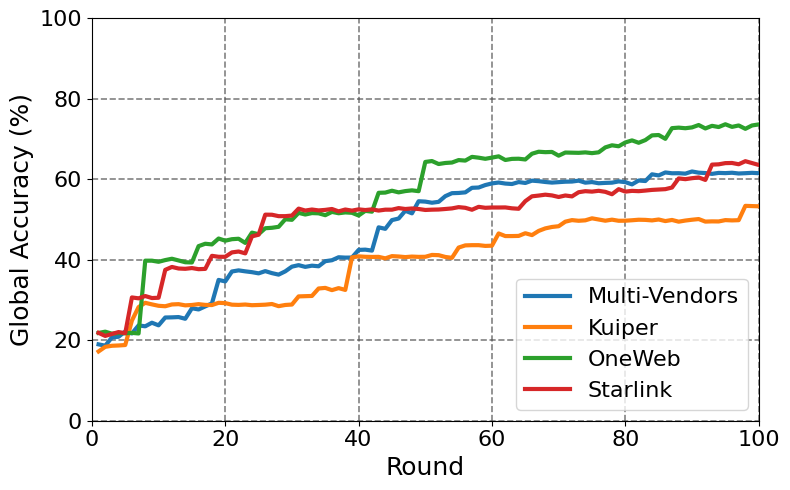} 
    \caption{20-minute slack time.}
  \end{subfigure}\hfill
  \begin{subfigure}[b]{0.32\textwidth}
    \centering
    \includegraphics[width=\linewidth]{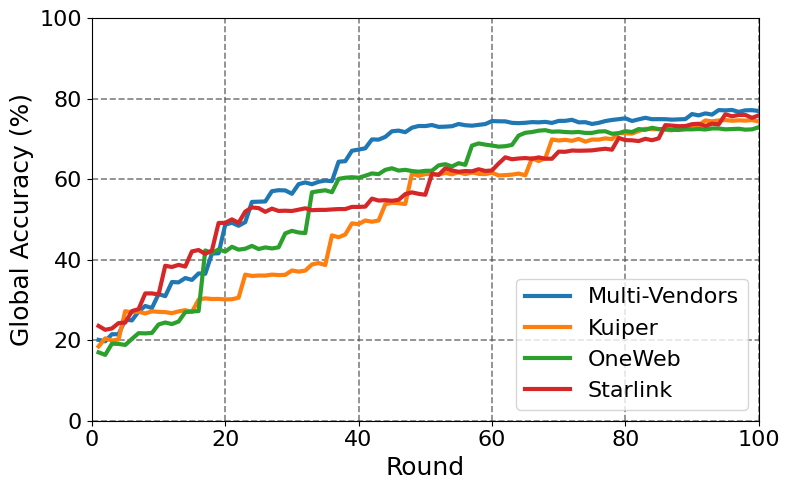} 
  \caption{30-minute slack time.}
  \end{subfigure}
\caption{\textsc{OrbitChain} convergence speed on the \textbf{EuroSat} dataset under different slack times (10, 20, and 30 minutes) and varying data heterogeneity levels. Subfigures (a)--(b) present results for the i.i.d.\ setting, while subfigures (c)--(d) show the corresponding non-i.i.d.\ performance.}
  \label{fig:Euro_Eval}
\end{figure*}

\vspace{2mm}
\noindent{\bf  PoA Consensus Performance and Latency Analysis.} Our results show a clear relationship between quorum size and the latency–stability trade-offs of the PoA consensus layer. As shown in \fref{fig:avg_latency}, average block finalization time increases with the quorum threshold: approximately $0.16$\,s for 1-of-5, $0.26$\,s for 3-of-5, and $0.35$\,s for 5-of-5. This growth reflects the expected coordination overhead of requiring more validator signatures per block.

Figures~\ref{fig:poa1}--\ref{fig:poa5} provide additional insight across 1{,}000 sequential block commits. The 1-of-5 configuration achieves the lowest latency but shows a gradual upward drift due to accumulated I/O overheads. The 3-of-5 configuration maintains a tighter and more stable latency distribution, demonstrating robustness to transient message delays while preserving sub-second responsiveness. In contrast, the 5-of-5 configuration yields the highest latency yet exhibits the smallest variance, indicating that full unanimity produces the most deterministic behavior once validators remain synchronized.

These results highlight the inherent PoA trade-off: smaller quorums offer faster responsiveness at the cost of weaker fault tolerance, whereas larger quorums improve consistency and accountability but introduce higher latency. In practice, the 3-of-5 configuration strikes the best balance for \textsc{OrbitChain}, achieving sub-second consensus while tolerating one Byzantine validator—an alignment well-suited to multi-vendor LEO environments. Even under elevated transaction loads (i.e., satellites' local models and aggregated global models by HAPs), the system sustained throughput above 200~tx/s, demonstrating that PoA can meet the real-time synchronization demands of FSL.

Overall, these findings confirm that \textsc{OrbitChain}'s PoA layer provides an effective balance of decentralization, determinism, and low latency, ensuring secure and auditable coordination among heterogeneous satellite vendors operating under real orbital communication constraints.

\vspace{2mm}
\noindent{\bf  Comparison to existing blockchain-based FSL baselines.} While prior work has explored blockchain-assisted FL for LEO constellations~\cite{pokhrel2021blockchain,wu2024sharded}, these systems differ fundamentally in both assumptions and architectural goals, making a direct performance comparison not methodologically meaningful. For example, Pokhrel \emph{et al.}~\cite{pokhrel2021blockchain} assume on-device proof-of-work (PoW) mining by satellites and UAVs—an approach that is computationally prohibitive for large-scale constellations and provides no mechanism for cross-vendor provenance, collusion resistance, or verifiable audit trails. Likewise, SBFL-LEO~\cite{wu2024sharded} operates within a single administrative domain and relies on satellites acting as miners/validators, focusing on CS+DBSCAN-based poisoning detection rather than long-term accountability, governance, or multi-vendor trust management. In contrast, \textsc{OrbitChain} is designed specifically {\em for federated, multi-vendor LEO ecosystems}: it offloads consensus to HAP-based PoA nodes, embeds immutable provenance metadata into every model update, and enforces accountability through quorum-backed signatures and vendor-scoped trust domains. None of these capabilities are present in existing blockchain–FL frameworks, nor do they address secure coordination among mutually distrustful satellite operators.

\subsection{\textsc{OrbitChain}'s Convergence Analysis Results}

This experiment evaluates the impact of enabling multi-vendor collaboration on \textsc{OrbitChain}'s convergence speed, without relying on any assumptions about intra- or inter-satellite link for model exchange—an ability that remains limited or unavailable for many commercial constellations and is often required by traditional FSL schemes to accelerate convergence. Instead, our analysis isolates the benefit that arises purely from aggregating heterogeneous local models across vendors. As shown in \fref{fig:MNIST_Conv}, we simulate three independent constellations, each operated by a different vendor. Each vendor’s satellites train locally on MNIST at varying slack times, during which the respective HAP operates in an idle-collection mode to receive as many model updates as possible from visible satellites. After each slack interval, HAPs across vendors jointly produce the global model for that round.

Under the i.i.d. setting with a 10-minute slack time, the multi-vendor global model exceeds 90\% accuracy within only four rounds (approximately 40 minutes). While some single-vendor configurations achieve similar performance—largely due to MNIST’s simplicity and the efficiency of CNN-based training—the differences become pronounced in the non-i.i.d. setting. Specifically, in the non-i.i.d. scenario, multi-vendor collaboration shows a clear advantage: the global model reaches over 85\% accuracy within 200 minutes and surpasses 95\% after 500 minutes. In contrast, single-vendor models converge significantly more slowly, requiring an additional \(\sim30\) minutes under a 10-minute slack time, and up to \(\sim200\) minutes more when slack time increases to 30 minutes. These results highlight that the heterogeneity and complementary coverage across vendors substantially accelerate convergence—especially under realistic, non-i.i.d. conditions.

\vspace{2mm}
\noindent{\bf Evaluation on real-world evaluation.}  In \fref{fig:UCMERCED_CONV}, we benchmark convergence on the UC Merced dataset. Under the i.i.d. setting with a 10-minute slack time, the multi-vendor global model reaches 80\% accuracy in ~570 minutes. In comparison, the fastest single vendor (Kuiper) reaches the same accuracy in 700 minutes, the second fastest (OneWeb) in ~920 minutes, while the third (Starlink) reaches only 20\% after 100 minutes—highlighting the stability and consistency benefits of multi-vendor aggregation. A similar trend appears with 20-minute slack time: the multi-vendor model reaches 80\% accuracy in ~600 minutes, whereas single-vendor models require 1,000–1,560 minutes. The results with 30-minute slack time show the same pattern.

Under the more challenging non-i.i.d. setting with a 10-minute slack time, the multi-vendor global model attains 80\% accuracy after 500 minutes, whereas some of the single-vendor models fail to reach this accuracy within 100 minutes, achieving only 57–62\%. Increasing the slack time to 30 minutes shows that the multi-vendor model can still reach 80\% accuracy in roughly 1,200 minutes, whereas some of the single-vendor models reach only 65–78\% even after 3,000 minutes—demonstrating that multi-vendor collaboration reduces convergence time by at least 2.5× while delivering higher or comparable accuracy. Importantly, increasing slack time primarily benefits {\em only} single-vendor learning, helping it catch up, whereas multi-vendor collaboration consistently converges faster and requires less slack time to reach the same or higher accuracy.

Finally, in \fref{fig:Euro_Eval}, we show the results on the EuroSat dataset. Consistent with the UC Merced findings, the multi-vendor scenario either matches or exceeds the performance of the best single-vendor models, and in several cases clearly outperforms them. This demonstrates the tangible benefits of cross-vendor collaboration. The alignment of these trends across two distinct remote-sensing datasets further underscores the robustness of the proposed multi-vendor collaboration and aggregation strategy.

\section{Conclusion}
We presented \textsc{OrbitChain}, a blockchain-backed framework for secure, auditable, and privacy-preserving multi-vendor FSL in LEO networks. By offloading consensus to HAPs and employing secure aggregation with age- and reputation-aware weighting, \textsc{OrbitChain} achieves robust convergence across heterogeneous satellites while protecting vendor data. Extensive experiments show up to 30 hours faster convergence on high-resolution satellite datasets compared to single-vendor FSL, with sub-second ledger finality and improved privacy, security, and global model accuracy. These results demonstrate \textsc{OrbitChain}’s practicality for trustworthy multi-vendor space-based AI collaboration.

{
\small
\bibliographystyle{IEEEtran}
\bibliography{biblio}}

\end{document}